\documentclass[fleqn,10pt]{wlscirep}
\usepackage{setspace}
\usepackage[utf8]{inputenc}
\usepackage[T1]{fontenc}
\usepackage{bm}
\usepackage{graphicx}
\graphicspath{ {./figures/}}
\def\vec{\mathbf}
\def\t#1{\text{#1}}
\usepackage[normalem]{ulem}

\title {Mathematical Operations and Equation Solving with Reconfigurable Metadevices}

\author[1]{Dimitrios C. Tzarouchis}
\author[2]{Mario Junior Mencagli}
\author[1]{Brian Edwards}
\author[1,*]{Nader Engheta}
\affil[1]{University of Pennsylvania, Department of Electrical and Systems Engineering, Philadelphia, PA, 19104, USA}
\affil[2]{University of North Carolina at Charlotte, Department of Electrical and Computer Engineering, Charlotte, NC, 28223, USA}
\affil[*]{corresponding author: engheta@@seas.upenn.edu}

\begin{abstract}
Performing analog computations with metastructures is an emerging wave-based paradigm for solving mathematical problems. For such devices, one major challenge is their reconfigurability, especially without the need for a priori mathematical computations or computationally-intensive optimization. Their equation-solving capabilities are applied only to matrices with special spectral (eigenvalue) distribution. Here we report the theory and design of wave-based metastructures using tunable elements capable of solving integral/differential equations in a fully-reconfigurable fashion. We consider two architectures: the Miller architecture, which requires the singular-value decomposition, and an alternative intuitive direct-complex-matrix (DCM) architecture introduced here, which does not require a priori mathematical decomposition. As examples, we demonstrate, using system-level simulations tools, the solutions of integral and differential equations. We then expand the matrix inverting capabilities of both architectures toward evaluating the generalized Moore--Penrose matrix inversion. Therefore, we provide evidence that metadevices can implement generalized matrix inversions and act as the basis for the gradient descent method for solutions to a wide variety of problems. Finally, a general upper bound of the solution convergence time reveals the rich potential that such metadevices can offer for stationary iterative schemes.
\end{abstract}

\begin{document}
\flushbottom
\doublespace 
\maketitle
\thispagestyle{empty}

\section*{Introduction} 

Naturally occurring or artificially designed systems are often modeled by linear equations, which can be expressed through simple algebraic entities, e.g., vectors and matrices. Building on these entities, numerical linear algebra describes the methods for performing a large variety of operations, such as vector-matrix multiplications and matrix inversions, which enable the extraction of valuable information regarding the system under study~\cite{Trefethen1997}. The most successful platform for numerically studying these systems is through the digital electronic computer. However, as these devices reach their limits with regards to speed and power-consumption considerations~\cite{Caulfield2010}, the idea of designing more efficient, application-specific, analog platforms has gained significant momentum in recent years~\cite{Solli2015}. To this end, one can find two major areas of hardware-enhanced analog computational approaches: electronic~\cite{Wang2020} and wave-enabled (free-space~\cite{Silva2014,Pors2015,Wetzstein2020} and waveguided~\cite{Harris2018,MohammadiEstakhri2019}) systems.

In the very mature electronic domain, there are examples of reconfigurable memristive devices that implement non-von Neumann/neuromorphic~\cite{LeGallo2018,Mead2020} crossbar architectures that perform enhanced analog real-valued matrix-vector operations~\cite{LeGallo2018} and solve real-valued mathematical equations~\cite{Zidan2018,Sun2019,Mead2020}. In the free-space optical domain, analog vector-matrix multiplication for image processing and deep-learning is performed through a cascade of lenses and volumetric scatterers~\cite{Vellekoop2007,Zhou2020,Wetzstein2020}. However, given conventional optical systems' inherent bulk, the flat-optics/metasurfaces demonstrate mathematical functionalities, e.g, computational imaging~\cite{Imani2020}, mathematical operations (integration/differentiation~\cite{Silva2014,Pors2015,DelHougne2018,Cordaro2019}), and discrete Fourier transformations~\cite{Matthes2019}, in a much more compact manner.  

In parallel, waveguided systems, such as integrated photonics, are proving to be a suitable platform for analog - classical~\cite{Feldmann2021} and quantum~\cite{Harris2018}- computations and machine learning tasks~\cite{Shen2017,Hughes2019}. These systems are designed using a combination of tunable and reconfigurable directional couplers, e.g., Mach--Zehnder interferometers (MZIs)~\cite{Haffner2015}, or tunable photonic waveguides~\cite{Wu2014,WuZ2014,Taballione2019,Bogaerts2020} and other wave-based tunable technologies, e.g., graphene and semiconducting nanopillars ~\cite{Low2014,Morea2018,Fan2019}. The majority of the MZI-based systems utilize the network designs introduced by: (a) Reck et al.~\cite{Reck1994} (first of its kind at the optical/quantum domain), (b) Miller~\cite{miller2013,Miller2015} (with \emph{self-configuring} characteristics), and (c) Clements et al.~\cite{Clements2016} (balanced path lengths). All three architectures require the a priori mathematical decomposition of the main operator. Like optics, photonic devices are orders of magnitude larger than the operation wavelength. Hence, inverse-designed metastructures are emerging examples for compact, ultrafast, low-power, and parallel analog computations~\cite{MohammadiEstakhri2019,Zangeneh-Nejad2020}. 

The majority of the aforementioned wave-based approaches, including the neural network approaches~\cite{harris2017}, deal with the matrix-vector product problem (wave interference). For the matrix inversion problem, one can find works based on optical systems~\cite{Psaltis1979,Cederquist1979,Wu2014,Akins1980} and metastructures~\cite{MohammadiEstakhri2019}; the latter exhibiting limited reconfigurability. All the inversion systems~\cite{Psaltis1979,Cederquist1979,Akins1980,MohammadiEstakhri2019} (including the ones at the electronic domain~\cite{Zidan2018,Sun2019}), implement a simple Jacobi iterative scheme~\cite{S.1995}, by a properly designed feedback loop. The scheme converges only for square matrices with eigenvalues inside the unit circle~\cite{Goodman1982}. Inevitably, such devices cannot be readily used for the inversion of general matrices, e.g., rectangular (minimization/regression problems), singular (inverse scattering/imaging problems), or simple square matrices with complex eigenvalues outside the unit circle. Therefore, the need for a wave-based system that (1) intuitively offers full reconfigurability without any a priori mathematical decomposition of the operator and (2) can perform matrix inversion for general matrices is apparent. Here we introduce such a system.

This work is structured and presented in two parts. The first part deals with the metadevice architecture, while the second part addresses the extension of its capability to invert general matrices. Specifically, first, we theoretically demonstrate (via system simulations) the matrix inversion capabilities of a structure that uses the Miller architecture (requires singular value decomposition (SVD)) for solving integral equation (IE) problems in the complex domain. Next, we introduce an alternative direct-complex-matrix (DCM) architecture that requires no a priori mathematical calculations and can be tuned intuitively. The DCM architecture can be regarded either as a generalized form of a tunable beamforming/beamsteering network or the wave-analog of a crossbar architecture with complex algebra abilities. The DCM architecture is then used to find the solutions of a set of differential equations (DE). Finally, we use both architectures to perform the generalized inversion of a complex-valued singular and a binary-valued rectangular matrix, i.e., implementing the generalized Moore--Penrose inverse. 

The second part of this work relates to the choice of the DE problems, highlighting that even well-studied DEs may yield matrices that cannot be inverted with a naively applied Jacobi iteration method. For these purposes, we rethink the algorithm, introducing a mathematical modification inspired by the definition of the pseudoinverse~\cite{Penrose1955}, which is essentially equivalent to the modified Richardson~\cite{Richardson1911,Landweber1951} iteration method, an equivalent form of gradient descent method (GDM)~\cite{ShewchukJonathan1994}. In other words, this modification enables the generalized inversion of any matrix, including singular and rectangular cases. We demonstrate this method utilizing both architectures, extending the previously reported results~\cite{MohammadiEstakhri2019}, and we extract theoretical upper bounds that quantify the possible hardware-induced enhancement to the convergence of GDM. Both architectures demonstrate that metadevices may enable compact, parallel, low-power, ultra-fast platforms for general numerical linear algebraic problems (matrix inversion and the GDM) and open avenues for metadevice-powered generic algebraic explorations. The proposed metadevice can be suitably aligned with open quantum algorithmic~\cite{Harrow2009} and quantum computing~\cite{harris2017} forefronts. 

\begin{figure}[h]
\centering
\includegraphics[width=0.8\textwidth, angle=0]{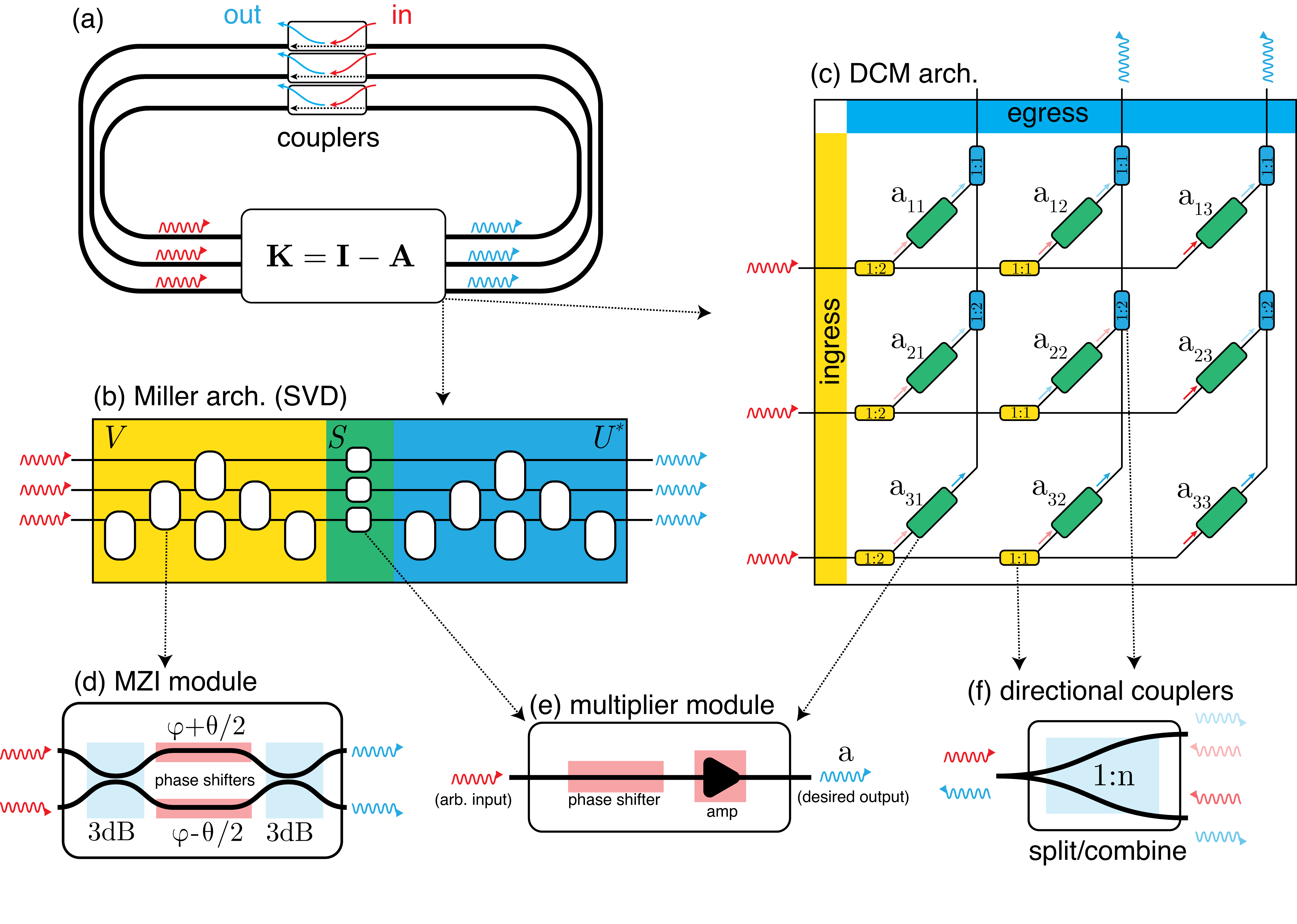}
\caption{\textbf{Reconfigurable Equation Solvers}:  Solving mathematical equations (or equivalently calculating the inversion of a matrix $\vec{A}$) requires the implementation of an operator (i.e., $\vec{K}=\vec{I}-\vec{A}$) within a looped system, schematically shown in panel (a). Here we explore two different reconfigurable architectures: (b) The Miller architecture consists of three stages, each representing the three matrices of the singular matrix decomposition. Stages $\vec{V}$ and $\vec{U}^*$ can be implemented as a cascade of Mach--Zehnder interferometers (MZI, panel (d)). In contrast, the $\vec{S}$ stage can be implemented either with MZI or with a multiplier module (panel (e)), where a given input signal can be modified accordingly via a phase shifter and an amplification/attenuation unit. (c) A more straightforward and more intuitive implementation of the same operator can be given in the introduced DCM architecture. Here, the ingress stage consists of directional couplers (panel (f)) with a fixed split ratio. Each of the outputs is driven to a dedicated multiplier (panel (e)) where the desired matrix value is assigned (here is an example of a $3\times3$ matrix). Finally, the signals are recombined at the egress stage using a set of directional couplers (panel (f)). Note that an MZI module can be reconstructed as two 3dB couplers and two multipliers, making the multiplier module the main constituent module for both architectures.}
\label{fig:1}
\end{figure}
\section*{Results}
\subsection*{Solving integral equations with the Miller architecture}
To begin with, we consider the following integral equation:
\begin{equation}\label{eq1_IE}
x\left( u \right) = {c(u)} + \int\limits_a^b {K\left( {u,v} \right)x\left( v \right)dv}
\end{equation}
which is known as the Fredholm integral equation of the second kind~\cite{Landweber1951}. Based on a feedback network consisting of $m$ input and $m$ output waveguides, Eq. (\ref{eq1_IE}) can be represented in matrix-vector form as ${\bf{x}}= {\bf{c}} + {\bf{K}}  {\bf{x}}$, where $\vec{x}\in\mathbb{C}^{m\times{1}}$ is the solution vector, ${\bf{c}}\in\mathbb{C}^{m\times 1}$ denotes the input vector, and ${\bf{K}}\in \mathbb{C}^{{m}\times{m}}$ is the kernel matrix obtained by sampling ${K\left( {u,v} \right)}$ in ${m}\times{m}$ points taking into account the appropriate $\Delta x = (b-a)/m$ factor.\footnote{Following the numerical algebra nomenclature, boldfaced capital and lowercase latin variables indicate matrices and vectors, respectively, while greek and unboldfaced latin letters indicate scalar quantities.} Although the previous vector equation is identical to the common feedback equation in electronics, its variable's physical meaning is fundamentally different. In electronics, the signals are represented as functions of time, and the processing occurs in the temporal domain; here, they represent the complex amplitude of monochromatic electromagnetic wave modes, and the processing occurs in the spatial domain \cite{MohammadiEstakhri2019,pozar}. The matrix ${\bf{K}}$ can then be interpreted as a transmission matrix relating a set of input waves to a set of output waves, thus describing the matrix-vector multiplication ${\bf{K}}   {\bf{x}}$. This linear operation can be performed in a variety of ways. Within this section, we employ an architecture based on the MZI mesh proposed by Miller \cite{miller2013,miller2017} (see Fig.~(\ref{fig:1})). Such a mesh configuration, exploiting the SVD, enables us to implement any ${m}\times{m}$ transmission matrix. In addition to the advantages of the previous metamaterial-based equation solver \cite{MohammadiEstakhri2019} (i.e. phase stability, low-power consumption, compatibility with silicon photonics technology~\cite{harris2017}, and implementation of more general, translational (shift) variant kernels), this approach presents two other features.  First, this platform is fully programmable. With high reconfigurability, it can perform multiple functions such as solving linear integral and differential equations, and performing matrix inversion. Second, contrary to inhomogeneous materials, an MZI mesh that implements a kernel operator can be trained by simple progressive algorithms \cite{miller2013} without involving global optimization techniques.

\begin{figure}[ht]
\centering
\includegraphics[width=1\textwidth]{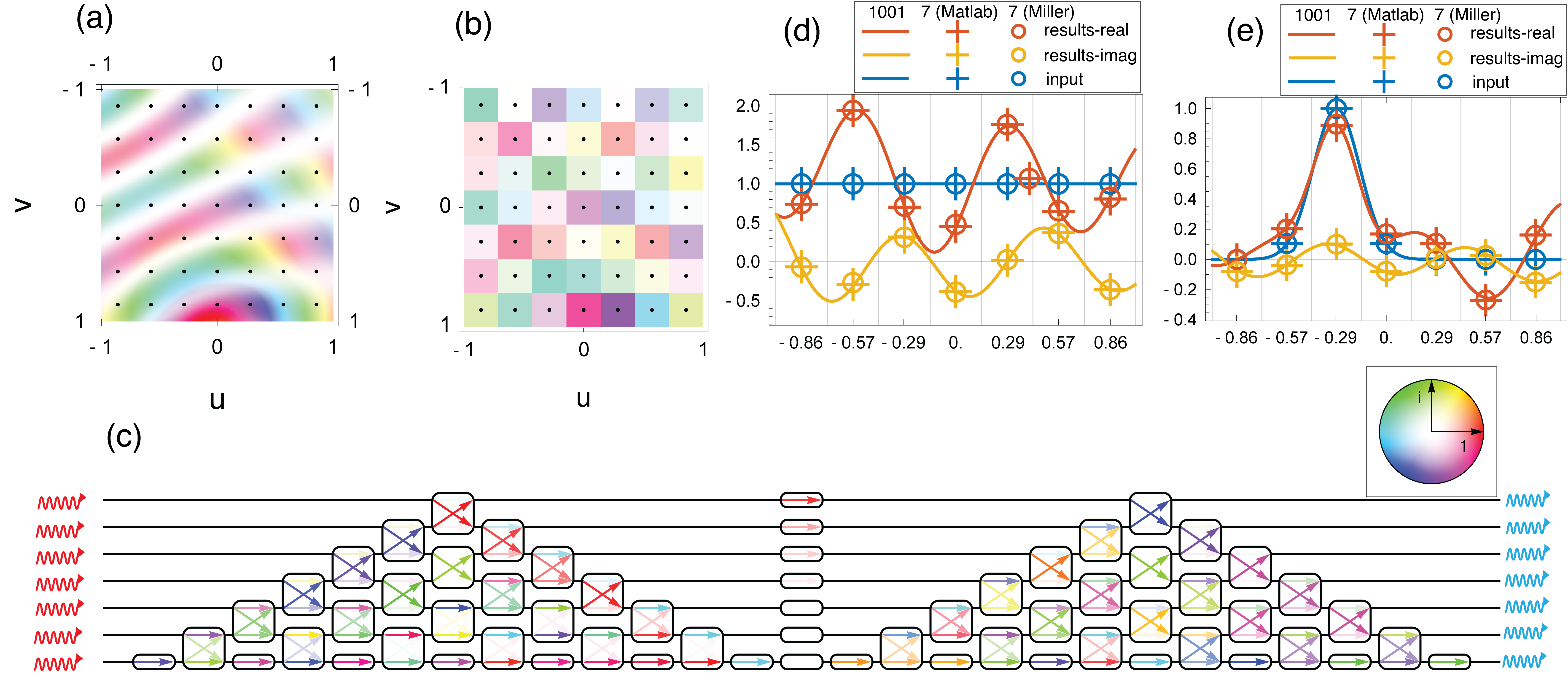}
\caption{\textbf{Solving Integral Equations with the Miller Architecture}: (a) Continuous distribution of the kernel $K(u,v)$. The black dots denote the spatial points where the kernel is sampled to derive its 7x7 discrete version $\bf{K}$ reported in the color plot (b). The maximum amplitude of the complex value in the color circle insets is set to 1.15 and (2/7)(1.15) in panels (a) and (b), respectively. (c) Schematic of flow through the MZI network wherein the color of the arrows denotes the complex transmission between the MZI ports with a maximum value of 1. (d\&e) The extracted solution (real-red and imaginary-yellow) of the integral equation (\ref{eq1_IE}) with the proposed kernel under two different real inputs (blue). In (d), the input is a constant value $1$, while in (e), it is an appropriately sampled Gaussian distribution centered on $u=-2/7$. This solution is evaluated in three different ways: the highly sampled (n=1001) \emph{true} solution (continuous line) evaluated using linear algebra (i.e. $\mathbf{x} = \left(\mathbf{I} - \mathbf{K}\right)^{-1}  \mathbf{c}$) method, the solution (crosses) with $n=7$ also calculated directly using linear algebra, and the solution (circles) calculated via the simulation of our MZI mesh.  All three show excellent agreement.}
\label{one_IE}
\end{figure}

To test the performances of the proposed MZI-based computing platform, as an example here we employ it to solve Eq. (\ref{eq1_IE}) characterized by the following kernel (Fig.~\ref{one_IE}(a))
\begin{equation}\label{eq2_IE}
K\left(u,v \right) = 1.2 H_0^{(2)}\left(8\sqrt{(u+1)^2+(v+1)^2}\right) + 1.7i~ H_0^{(2)}\left(8\sqrt{u^2+(v-1)^2}\right)
\end{equation}
\textcolor{black}{where $H_0^{(2)}(.)$ is the Hankel function of the second kind and order 0}, defined over $\left[ {a,b} \right] = \left[ { - 1,1} \right]$. Considering a feedback network composed of $m=7$ waveguides, the ${7}\times{7}$ transmission matrix ${\bf{K}}$ is obtained based on the kernel distribution (Fig.~\ref{one_IE}(b)). 
Using the SVD technique, ${\bf{K}}$ is then written as the product of three transmission matrices: ${\bf{K}}={\bf{V}}  {\bf{S}} {{\bf{U}}^* }$, where $*$ denotes the complex conjugate transpose operation. {\bf{U}} and {\bf{V}} are unitary transmission matrices and are respectively implemented by the left and right triangular-shaped MZI sets of Fig.~\ref{one_IE}(e). These MZI sets have been configured by a compiled programming protocol described in \cite{miller2013}, whose phase shifters settings for each MZI result from a progressive algorithm based on unitary operator factorization. {\bf{S}} is a ${7}\times{7}$ diagonal transmission matrix of nonnegative real numbers smaller than one satisfying the convergence condition \cite{MohammadiEstakhri2019} and has been implemented by $7$ multipliers connecting the two triangular-shaped MZI sets (see Fig.~\ref{one_IE}(b)). These multipliers have been configured to match the diagonal elements of $\bf{S}$.

   It is worth noticing that the proposed computing platform can also be applied to kernels that do not satisfy the convergence condition. As indicated in literature \cite{MohammadiEstakhri2019}, in such cases, the kernel and the input signal can be properly scaled down to meet the convergence condition and ensure that integral equation recursive and exact solution remains proportional. Satisfying the convergence condition in a wave-based recursive computing platform implies that any time the waves circulate through the structure representing the kernel, they lose part of its energy. Unlike the approach utilized in \cite{MohammadiEstakhri2019}, where such energy gets dissipated by absorbing materials, in the platform here, that energy is not lost and is still available at non-connected ports of MZIs implementing $\bf{S}$.  As a result, it can be potentially used for other purposes.

The MZI structure, the waveguide feedback network, and in/out couplers with $1{\%}$ coupling coefficient were numerically simulated with a system-level software~\cite{AWR,supp}. Figs.~\ref{one_IE}(c) and \ref{one_IE}(d) present the solutions (real and imaginary part) of Eq.~(\ref{eq1_IE}) with the kernel in (\ref{eq2_IE}) for two different inputs, obtained using (i) the proposed approach using MZIs (shown as dots) in which the discrete solution $\bf{x}$ has been extracted from the couplers in the feedback network, (ii) a standard numerical inversion of the corresponding 7x7 linear system (shown as crosses), and (iii) theoretical approach, i.e., the linear system resulting from the oversampling of the kernel approaching its continuous variation over the entire chosen spatial domain (shown as a solid line). As can be seen, the MZI-based structure's solutions agree well with the corresponding linear equations' exact solutions and the exact almost continuous solutions derived by oversampling the kernel~\cite{supp}.

\subsection*{Solving differential equations with the DCM architecture}

The adaptation of the Miller architecture for a metastructure that is able to solve integral equations introduces a compact, elegant and self-configuring realization for any given operator,  provided a preliminary SVD decomposition of the operator itself is performed beforehand. However, its elegance comes at a cost, since performing the SVD for a matrix is as computationally involved as inverting the same matrix. In the context of solving a set of linear equations (i.e. matrix inversion), preferably one might desire to implement and solve any kernel without incurring this cost. This section describes the DCM network architecture that allows the realization of any matrix in a simple, intuitive, and direct manner, without the need for any a priori calculations. 

As seen in Fig.~\ref{fig:1} the alternative architecture consists of three stages: the \emph{ingress} (yellow), \emph{middle} (green), and \emph{egress} (blue) stages. The input signal vector is directed at the ingress stage, where each one of the $m$ inputs is split into $m$ outputs to yield $m^2$ distinct values ($m$ lines, split $m$ times), each of them with relative amplitude $1/\sqrt{m}$. 

The ingress stage can be implemented in a multitude of ways. For example, one could use either $m$ signal-splitters (using inverse-designed waveguides or RF lumped elements, etc.) or $m$ cascades of $m-1$ properly connected directional couplers or MZIs. In either case, these devices are fixed and are not dependent on the choice of the kernel. Note that any amplitude or phase change that the ingress stage introduces can be compensated by the proper adjustment of the components at the middle stage.

The $m^2$ outputs of the ingress stage are directed into $m^2$ dedicated \emph{complex multipliers}, i.e., the middle stage. An individual complex multiplier changes the amplitude and phase of its input signal, and is a direct implementation of an element of the kernel matrix. A possible implementation of a multiplier module can be achieved either by using a phase-shifting component followed by an amplifier (that offers a dynamical range of amplification or attenuation), or a tunable directional-coupling module, or an MZI. The first choice might be particularly interesting and feasible in the RF-electronics platforms while the MZI approach can be particularly attractive for integrated photonics platforms. 

Finally, the $m^2$ outputs of multipliers are directed to the final egress stage. This stage operates similarly to the ingress but in the reverse direction. For example, it combines $m^2$ input signals into $m$ output signals (as the output vector), which are then directed to the feedback-coupling modules. The $m$ outputs are consequently directed to the feedback loop stage, that essentially enables the extraction of the main matrix inversion. \footnote{\textcolor{black}{It is important to note the egress stages is not a power combiner. The output amplitude of each of the $m$ egress parts is proportional to the fraction (i.e., $1/\sqrt{m}$) of the algebraic sum of complex-valued amplitudes of $m$ input signals to it, and consequently the output power is usually less than the sum of the input power. As a result, the remaining power goes out of the system in unused channels in the MZI mesh in egress. After the egress stage we need to have amplifiers to bring the output signal level up to compensate for the $1/\sqrt{m}$ of the ingress and the $1/\sqrt{m}$ of the egress (i.e., therefore compensate for $1/m$).}}

\textcolor{black}{The DCM architecture functionally shares a few similarities with the electronic crossbar architectures, e.g., as in~\cite{Zidan2018, Sun2019}, only this time using waveguiding systems. The electronic crossbar (or crosspoint) architecture utilizes Ohm's and Kirchhoff's law combined with analog and reconfigurable resistive memories to perform real-valued calculations. In our case, the reconfigurable resistive memories are the multiplier modules that enable full-complex arithmetic manipulation. At the same time, the ingress/egress stages utilize the wave signal splitting and recombining to deliver the signal towards all the multiplier modules and the output. In contrast, the electronic counterparts utilize Ohm's and Kirchhoff's law (voltage division and current-voltage proportionality) for the same purpose.}.

\begin{figure}[h]
\centering
\includegraphics[width=1\textwidth]{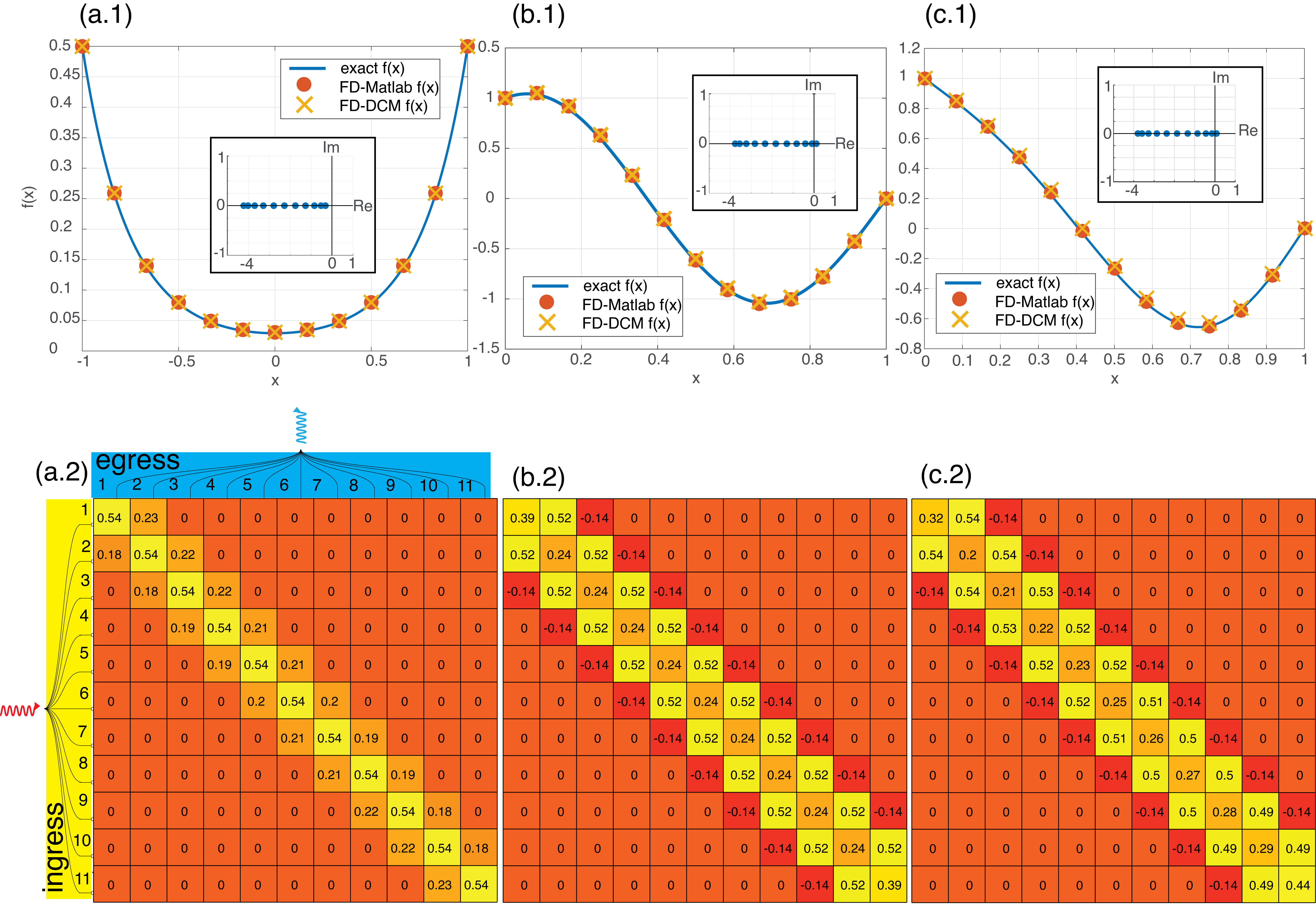}
\caption{\textbf{Solving Differential Equations with the DCM Architecture}. A schematic representation of an $11\times11$ network is depicted, with the corresponding element numbering. (a.1-a.2) The result for the case of a Hermite equation in the interval $x\in[-1,1]$ with boundary conditions $f(-1)=f(1)=1/2$. The inset figure depicts the distribution of the eigenvalues of the corresponding difference matrix ${\bf A}$ in the complex plane. For this case, all eigenvalues are negative (reside in the left-half plane). The corresponding color-coded matrix in (a.2) represents the values for each of the multipliers of the alternative network; (b.1-b.2) Similarly, the solution of a Helmholtz equation in the range $x\in[0,1]$ with $f(0)=1$ and $f(1)=0$. For this case, one of the eigenvalues of the difference matrix resides in the right-half plane; (c.1-c.2) Finally, the solution of an Airy equation for $x\in[0,1]$, $f(0)=1$ and $f(1)=0$. For case (a) the kernel matrix is derived using the following formula $\vec{K}=\vec{I}-\alpha_\lambda \vec{A}$, where $\alpha_\lambda$ is just a scaling factor, while for the cases (b) and (c) the implemented kernel was derived by $\vec{K}=\vec{I}-\alpha_\lambda\vec{A}^*\vec{A}$.} 
\label{fig:3}
\end{figure}

Although the proposed alternative architecture can be applied to solve IEs, here we apply it to solve another major class of problems, ordinary differential equations, even with non-constant coefficients.
As we mentioned above, the existence of a feedback loop with a certain initial excitation $\bf{c}$ along the main kernel $\bf{K}=\bf{I}-\bf{A}$ yields to the matrix inversion of the following problem $\vec{A}=\vec{I}-\vec{K}$, hence the solution of any linear set of equations can be obtained. The key for solving differential equations (DE) is the proper discretization and the formulation of the equivalent linear problem, akin to the case of integral equations. 

For reference, the 1D (in terms of $x$) DE problem of 2nd order can be generally expressed as
\begin{equation}
\mathcal{D}:~
  \begin{cases}
    r(x)\frac{d^2f(x)}{dx^2}+p(x)\frac{df(x)}{dx}+q(x)f(x)=g(x), & x\in[x_\t{d},x_\t{u}]\\
    f(x_\t{d})=a,~f(x_\t{u})=b
  \end{cases}
\end{equation}
where the functions $r(x),p(x),q(x),$ and $g(x)$ are general complex functions while the values $a,b$ are the boundary conditions at the extreme points of a given interval $[x_\t{d},x_\t{u}]$ for which the solution of the differential equation is desired. While the above expression implements a Dirichlet-type boundary conditions, other types of boundary conditions can also be implemented in a similar manner. 

Following a standard Finite Differences (FD) method (see Methods section) 
one can obtain an equivalent difference equation ($\Delta$E), i.e., a discrete version of DE, whose solution (inversion) yields the continuous solution of the DE providing the mesh is sufficiently dense. The corresponding discrete version of the problem reads
\begin{equation}
\mathcal{D}_\Delta:~
  \begin{cases}
    r(x_j)\left(f_{j+1}-2f_j+f_{j-1}\right)+p(x_j)\frac{f_{j+1}-f_{j-1}}{2}(\Delta x)+q(x_{j})f_j(\Delta x)^2=g(x_j)(\Delta x)^2, & j\in\{1,...,m\}\\
    f_0=a, f_{m+1}=b
  \end{cases}
\end{equation}
where $x$ is the unknown variable and $\Delta x=\frac{|x_\t{d}-x_\t{u}|}{m}$ is the difference constant of the interval $x\in[x_\t{d},x_\t{u}]$ discretized in $m$ pieces. In this way it is possible to express the problem as $\vec{A}\vec{x}=\vec{c}$, where $\vec{A}\in\mathbb{C}^{m\times m}$ is the difference matrix, $\vec{c}\in\mathbb{C}^{m\times1}$ is the input vector (including the boundary conditions) and $\vec{x}\in\mathbb{C}^{m\times 1}$ is the unknown solution vector. 
As before, the solution vector $\vec{x}$ can be found once the kernel of the metadevice reads $\vec{K}=\vec{I}-\vec{A}$. It is worth noticing that the matrix inversion is affected by the eigenvalues of the kernel, i.e., the eigenvalues should be strictly within the unit circle in order for the iterative method to converge. The behavior of the kernel follows the spectral behavior of $\vec{A}$. In general~\cite{supp}, 
the combination of the arbitrary complex functions $r(x),p(x),q(x)$ with $\vec{A}_{\Delta x}$ (diagonally dominant~\cite{supp2}), $\vec{P}_{\Delta x}$ (skew symmetric) and $\vec{I}$ (unit matrix) can lead to $\vec{A}$ matrices with arbitrarily distributed complex eigenvalues. The above point can affect the convergence of the implemented iterative method. Previous reports on solutions of integral and differential equations with hardware-accelerated approaches~\cite{Zidan2018,Sun2019,MohammadiEstakhri2019} demonstrated the implementation of a matrix kernel with eigenvalues within the unit circle, i.e., a condition that guarantees the convergence of the iteration method. It may appear that this feature limits the applicability of these devices for solving more general problems. Here, however, we address this problem by carefully preconditioning the kernel and modifying the input, which eventually can lead to the realization that the proposed hardware-accelerated or hardware-implemented solvers, including our proposed architectures, can expand the range of their applicability towards a larger problem set.

Let us first extract and examine the peculiarities of certain well-known DE equations. For these examples we considered an $11\times11$ network consisting of ideal directional couplers and MZIs using the AWR Microwave Office\textregistered~\cite{AWR}. Similar to the previous section, here the results are obtained using $1\%$ (-20dB) couplers for the feedback loop~\cite{supp}. Additionally, all the results are compared against the exact solution and a FD code that implements the same method in MATLAB (see Methods). As the first example, we start with the Hermite equation that reads 
\begin{equation}
\frac{d^2}{d x^2}f(x)-2x\frac{d}{d x}f(x)+k f(x)=0
\end{equation}
This particular equation is useful in a multitude of fields since it is the basis of the Hermite polynomials~\cite{abramowitz1964handbook}. The general well-known solution reads 
\begin{equation}
  f(x)=c_1\t{H}_{\frac{k}{2}}(x)+c_2~_1\t{F}_1(-\frac{k}{2},\frac{1}{2},x^2)
\end{equation}
where $\t{H}_{n}(x)$ is the Hermite polynomial and $_1\t{F}_1(x)(a,b,x)$ is the confluent hypergeometric function of the first kind~\cite{ErwinKreyszig2011}. In our case we seek for a solution on the interval $x\in[-1,1]$ with conditions $f(-1)=f(1)=\frac{1}{2}$. Following the above analysis we have $r(x)=1,~p(x)=-2x,~q(x)=-10$, and $g(x)=0$. The solution of the above is $f(x)=\frac{3}{38 e}e^{x^2}\left(1+4x^2+\frac{4}{3}x^4\right)$. For this particular case the matrix $\vec{A}$ has eigenvalues with a negative real part. The "modified" kernel matrix $\vec{K}=\vec{I}-\alpha_\lambda\vec{A}$, where the additional scale factor is $\alpha_\lambda=-0.2$ such that all eigenvalues of the modified kernel lie within the unit circle. Eigenvalues bound theorems, such as Gershgorin's theorem~\cite{ErwinKreyszig2011}, can provide a good estimation of how large the scaling factor should be. As we can see in Fig.~\ref{fig:3}~(a.1), a comparison between the analytical solution, the simulated results from the MZI implementation and the corresponding MATLAB code reveals an excellent agreement. A denser discrete mesh of the interval (larger matrix implementation) can further improve the results' accuracy. An interesting feature here is that for these types of problems where the inverse of sparse matrices is required, one could avoid the implementation of a full $m\times m$ network, but rather a much smaller subset (see for example Fig.~\ref{fig:3}~(a.2)).  

The second example examines the solution of Helmholtz's equation, i.e., 
\begin{equation}
\frac{d}{d x^2}f(x)+k^2f(x)=0
\end{equation}
on $x\in [0,1]$ with $f(0)=1$ and $f(1)=0$. \textcolor{black}{For this particular example we assume $k=5$. In this case the solution reads $f(x)=c_1e^{ikx}+c_2e^{-ikx}$ with $c_1=\frac{1}{1-e^{10i}}$ and $c_2=\frac{-e^{10i}}{1-e^{10i}}$. }

Although this is a seemingly simple equation we observe that with a proper choice of $k$ and $\Delta x$ the difference matrix $\vec{A}$ can have eigenvalues with real parts both positive and negative (inset plot at Fig.~\ref{fig:3} (b.1)). For this case the direct implementation of such kernel will not converge. We can overcome this obstacle by implementing the following modified kernel instead
\begin{equation}
    \vec{K}=\vec{I}-\alpha_\lambda\vec{A}^*\vec{A}
\end{equation}
where $\alpha_\lambda=0.1415$ is a properly chosen scaling factor. In this case we can extract the solution once we apply the following input vector, $\vec{c}_\t{new}=\alpha_\lambda \vec{A}^*\vec{b}$. As we can see in Fig.~\ref{fig:3} (b.1) the results yield a very good agreement with the analytical curve, while the implemented matrix (Fig.~\ref{fig:3} (b.1)) shows that we have the implementation of an almost pentadiagonal matrix. Note that the zero-valued components could be as well omitted, allowing the implementation of such a system with fewer MZIs/multipliers.   

Finally, we examine the solution of the Airy equation, i.e., 
\begin{equation}
    \frac{d^2}{dx^2}f(x)+k^2 x f(x)=0
\end{equation}
This is a function with non-constant coefficients with an analytical solution that is expressed in terms of the Airy functions of the 1st and 2nd order, i.e. 
\begin{equation}
 f(x)=c_1\t{Ai}\left(-\frac{k^2}{k^{4/3}}x\right)+c_2\t{Bi}\left(-\frac{k^2}{k^{4/3}}x\right)   
\end{equation}
where $c_1$ and $c_2$ are arbitrary constants~\cite{abramowitz1964handbook}. Here we examine the solution for $k=2\pi$ in the range $x\in[0,1]$ with $f(0)=1$ and $f(1)=0$ ($c_1=1.16+i0.42$ and $c_2=0.95-i0.24$). As before, the difference matrix exhibits some eigenvalues with negative and positive real parts (inset plot at Fig.~\ref{fig:3} (c.1)), hence the utilized preconditioned is with the scale factor $\alpha_\lambda=0.138$. As we can observe in Fig.~\ref{fig:3}~(c.1) the results have an excellent agreement. Notice that both Fig.~\ref{fig:3}~(b.2) and Fig.~\ref{fig:3} (c.2) cases implement a quite similar (not identical) kernel.   

The results for all the above cases demonstrate several notable features. First, the kernel has been intuitively implemented via the DCM architecture, yielding results with excellent agreement compared to the analytical solutions. Second, the proposed spatially discretized approach allows for the solution of general differential equations, especially those with non-constant coefficients, such as the Hermite and Airy equations. This is a feature that clearly outperforms other conventional hardware-based approaches that use a Fourier transformation approach in the time domain~\cite{Zangeneh-Nejad2019} - in these cases the solution of non-constant differential equations is complicated. 
Third, some of the available solutions are expressed in terms of highly specialized forms, such as the Airy, Hermite, and Hypergeometric functions. In all these cases the MZI network actually generates the expected results, demonstrating that the proposed machine can be used as an alternative way for the evaluation of cumbersome/generalized functions. As a reminder, the used methodology is subject to all mathematical restrictions of the FD methodology~\cite{leveque2007finite}.  

\subsection*{Wave-enhanced GDM and its convergence bounds}

Let us first examine some of the details and implications of the preconditioning that we described above. In all the works where the solution of equations is desired, essentially the following basic iterative scheme
\begin{equation}\label{eq:iter1}
\vec{x}_{n+1}=\vec{c}+\left(\vec{I}-\vec{A}\right)\vec{x}_{n}
\end{equation} is utilized, where $\vec{A}$ is the desired square operator/matrix to be inverted and $n$ is the iteration number. This scheme can indeed lead to a solution for the linear problem when the unknown vector $\vec{x}$ has sufficiently close values for two iterations, i.e., $||\vec{x}_{n+1}-\vec{x}_{n}||<\epsilon$, where $\epsilon$ is the desired accuracy error. Once this criterion is achieved, then we may conclude that  $\vec{x}_\t{n+1}\approx\vec{x}_\t{n} \approx \vec{x}=\vec{A}^{-1}\vec{c}$. 
Despite the conceptual simplicity of Eq.~(\ref{eq:iter1}), the convergence of the method depends entirely on the distribution of the (complex) eigenvalues of the kernel $\vec{K}=\vec{I}-\vec{A}$, i.e., the eigenvalues should reside within the complex unit circle.~\cite{Goodman1982, Akins1980}. Moreover, all the recent examples, e.g., with inverse-designed metastructures~\cite{MohammadiEstakhri2019} or with electronic memristor array technology~\cite{Zidan2018,Sun2019}, demonstrate the inversion of well-behaved (non-singular) square matrices that satisfy the aforementioned convergence criterion.  

From the solution of DE it became apparent that even well-defined problems can yield matrices with eigenvalues beyond the unit circle, making the convergence of the implemented feedback scheme challenging. While for some cases a simple scaling of the kernel suffices, we examined some examples where the solution is possible only after the proper preconditioning of both the kernel and the input of the system. Here we examine that the $\vec{A^*A}$ modification, introduced in for optical systems in~\cite{Goodman1982}, can have much wider implications that enable the utilization of electronic or wave-based systems for inverting a much larger set of matrices. Actually, the aforementioned preconditioning allows us to evaluate the inverse of square matrices with arbitrary eigenvalues, but also to implement the notion of the generalized inverse (a.k.a. pseudoinverse) for rectangular and singular matrices, effectively introducing the GDM for the proposed metadevice architecture. 

In full form, this simple modification reads 
\begin{equation}\label{eq:iter2}
\vec{x}_{n+1}=\alpha_\lambda\vec{A^*c}+\left(\vec{I}-\alpha_\lambda\vec{A^*A}\right)\vec{x}_n
\end{equation}
which is nothing but a version of the well-known \emph{Richardson iteration}~\cite{Richardson1911}, implementing essentially the geometric series version for operators (Neumann series). This iteration has been rediscovered in several contexts and forms, such as the \emph{Landweber} iteration~\cite{Landweber1951}, and it represents a special form of the GDM~\cite{S.1995}. 

At this point, a couple of observations can be made for the above expression. First, the quantity $\vec{A^*A}$ forms a positive semidefinite matrix for any rectangular matrix $\vec{A}$, hence all of its eigenvalues reside in the positive real plane. Moreover, the factor $\alpha_\lambda$ can scale all the positive eigenvalues of $\vec{A^*A}$, hence the new modified kernel $\vec{K}=\vec{I}-\alpha_\lambda\vec{A^*A}$ always satisfies the convergence condition. As before, when $\vec{x}_{n+1}\approx\vec{x}_{n}$ we have that the iteration yields to the solution $\vec{x}=(\vec{A^*A})^{-1}\vec{A^*}\vec{b}$. This approach coincides with the definition of the generalized inverse or pseudoinverse of a matrix~\cite{Penrose1955}. Therefore, this simple modification enables the inversion of any given matrix. Additionally, GDM is a simple yet robust method for the evaluation of linearized problems, least-square problems, pseudoinverses, and unconstrained minimization problems~\cite{Trefethen1997}. Therefore, the proposed wave-based metadevice can in theory perform calculations and explorations for a much wider numerical linear algebra domain. However, GDM is a stationary method since both the step (scaling) factor $\alpha_\lambda$ and the kernel remain constant (stationary) at every iteration~\cite{S.1995}. The possibility of expanding the proposed devices toward non-stationary numerical methods is an obvious next step directly affected by the reconfigurability and the output readout speed. 

\begin{figure}[h]
\centering
\includegraphics[width=0.85\textwidth]{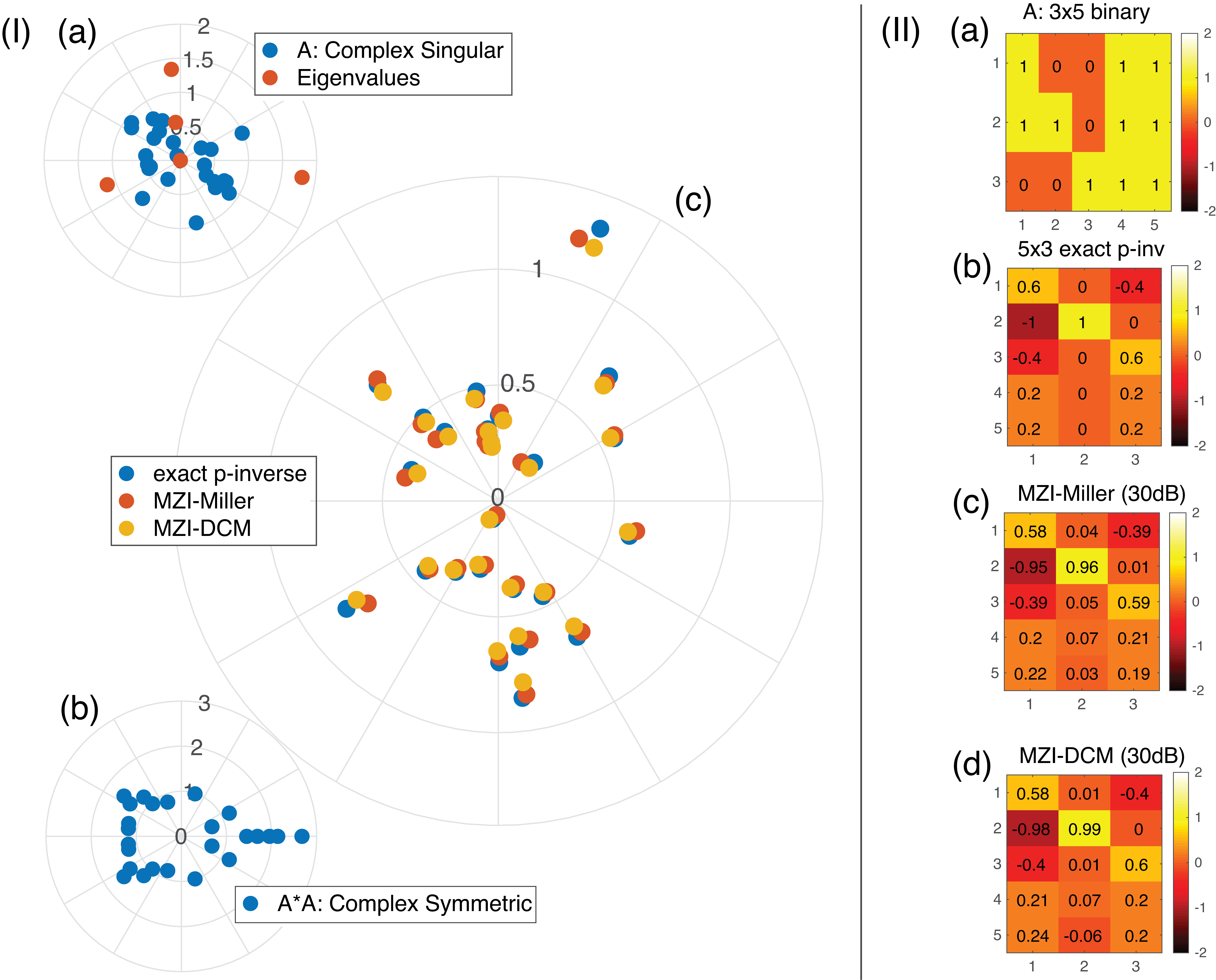}
\caption{\textbf{Generalized Inversion (Pseudo-Inversion) of Singular and Rectangular Matrices Using the Miller and DCM Architectures}. (Column I) The entries (blue dots) and the eigenvalues (red dots) of a singular $\vec{A}\in\mathbb{C}^{5\times5}$ matrix ((a) polar plot) and the entries of the preconditioned complex symmetric matrix $\vec{A}^*\vec{A}$ ((b) polar plot). The central polar plot (c) depicts the entries of the pseudoinverse $\vec{A}^+$ (blue dots) compared against the results obtained via the Miller (red dots) and the DCM network architecture, both utilizing the optimum scaling factor $\alpha_\lambda=\frac{2}{\lambda_1+\lambda_5}\approx0.316$. (Column II) The entries of a binary $\vec{A}\in\mathbb{Z}^{3\times5}$ matrix (a) and the corresponding pseudoinverse (b). The bottom figures depict the estimated pseudoinverses via the Miller (c) and the DCM network architecture (d), for $\alpha_\lambda=0.125$ and a feedback coupler with a 30dB coupling ratio.}
\label{fig:4}
\end{figure}

In support of the above theoretical observations, in Fig.~\ref{fig:4} we demonstrate the pseudoinverse of a singular $\vec{A}\in\mathbb{C}^{5\times5}$ and a rectangular binary $\vec{A}\in\mathbb{Z}^{3\times5}$ matrix, utilizing both the Miller and the DCM architecture. For the first case (Fig.~\ref{fig:4} (I)) the pseudoinverse of a singular matrix (one eigenvalue is zero - see inset Fig.~(~\ref{fig:4}) (a)) using the preconditioning technique is presented for both architectures. We observe an excellent agreement with the \emph{exact} solution (via MATLAB) for both platforms. 
For the second case (Fig.~\ref{fig:4} (II)), we observe that by applying the same preconditioning one can get an excellent estimation of the required generalized inverse of a binary rectangular matrix. Overall, the simulated results reveal that both architectures can successfully perform the required pseudoinverse calculation. 

Finally, the implementation of GDM (even in its simplest form) allows the extraction of a convergence bound that depends on both the properties of the matrix (eigenvalues) and the iteration timing (hardware). Assuming the case of a positive semidefinite matrix $\vec{A}$, the convergence of iteration in Eq.~(\ref{eq:iter1}) is of the order
\begin{equation}\label{eq:bound}
    \mathcal{O}\left(t\kappa(\vec{A})\log(\frac{1}{\epsilon})\right)
\end{equation}
where $\kappa(\vec{A})$ is the condition number of the matrix and $\epsilon$ is the desired approximation error (see Methods section). 

For either of the suggested wave-operated architectures, each iteration requires a total of $t=L\frac{\lambda}{c}$ time to complete, where $c$ is the speed of light, $\lambda$ the operating wavelength, and $L$ is the overall network length (dimensionless quantity, i.e., expressed in wavelengths). The overall network length varies depending the waveguiding systems that are used for either the Miller or the DCM. Assuming photonic implementations the network length can be 2 or 3 orders of magnitude larger than the wavelength (a single MZI operating at 1550 nm has a typical physical length at several $\mu$m scale). \textcolor{black}{A possible reduction of this length can be achieved with emerging ultracompact dielectric metamaterial approaches~\cite{Jahani2018,Mia2020,Song2022}, while novel inverse-designed microwave/photonic implementations can reduce the required length to a few wavelengths~\cite{MohammadiEstakhri2019}.} Moreover, the proposed metadevice can be implemented with lumped RF elements/technology, hence pushing the overall size to subwavelength regimes. Such size reduction could potentially lead to iteration timings at and below the nano/pico seconds range ($t\leq10^{-9}-10^{-12}$s). Therefore, the proposed metadevice holds the promise of an excellent playground for ultrafast numerical linear algebra techniques, within a given technological domain, such as optics, RF-circuits, photonics, and electronics. 

\section*{Discussion}
In conclusion, in this work, we theoretically demonstrated the matrix inversion capabilities of a structure that utilizes the Miller architecture and the introduced DCM architecture for solving IE and DE problems in the complex domain. In particular, the DCM architecture (the wave-analogous of the crossbar architecture in electronics) can intuitively implement any required matrix, without the need for the \emph{a priori} calculations required by the Miller architecture. 

We then performed the inversion (i.e., Moore--Penrose inverse) of complex-valued singular and binary-valued rectangular matrices using both architectures. The introduced mathematical modifications required for performing the above tasks revealed that the proposed device could perform the equivalent form of the gradient descent method. Finally, where possible, we extracted the upper convergence bounds of the inversion method. Those bounds depend on both the iteration time (platform dependent), the condition number of the matrix, and the required numerical accuracy of the computation. This bound revealed that matrix inversion could have the potential of ultrafast enhancement with the proper platform choice. However, the proposed metadevice and the presented methodology are platform-independent, offering a fertile path for RF and photonic experimental implementation. Therefore, the ideas presented here can further investigate compact, parallel, ultra-fast platforms for solving general numerical linear algebraic problems and other hardware-enhanced generic numerical linear algebraic methods with metadevices. \textcolor{black}{Currently, we are in the midst of an experimental investigation of our DCM architecture for matrix inversions and other advanced mathematical functionalities using an RF implementation.} 

\section*{Methods and Materials}
\subsection*{System simulations}
All presented results were obtained using AWR~\textregistered~\cite{AWR}Microwave Office, a commercially available RF circuital program able to perform system level simulations. For both architectures an MZI equivalent module was used as the main module. In the first case a $7\times7$ Miller matrix was constructed, while for the second case an $11\times11$ DCM architecture. The final results regarding the inversion of singular and rectangular cases were simulated in two $5\times5$ system for both Miller and DCM architecture~\cite{supp}.  

As for the number of simulated elements, the Miller architecture requires a total of $m^2+2m$ MZIs~\cite{Harris2018}, while the proposed DCM require a total of $m^2$ for the middle stage and $2m(m-1)$ for both the ingress/egress stage MZIs, assuming an MZI-only implementation. However, for DCM both the ingress and egress stages could as well be designed with fixed couplers/power splitting waveguiding systems, therefore the actual number of "active" (i.e., reconfigurable) elements is $\approx m^2$, comparable with the Miller approach.

Note that in our implementation each MZI requires the existence of 2 “active” tunable components (one for each waveguiding “arm” or one that controls tha amplitude and one for the phase of the output signal). In this case the Miller architecture requires $2m^2+4m$ active components. However, with the DCM with the multiplier requires two active elements (one phase shifter and one amplifier/attenuator unit), therefore it could be implemented with $2m^2$ active components. Therefore, any scalablity/insertion loss or other implementation related issues are practically dictated and limited by the platform used.

Concerning the number of waveguides $m$ used for representing the problem, it highly depends on mainly two factors: (a) the choice of the technological platform and the related experimental challenges and (b) the choice of the problem, e.g., kernels in solving integral/differential equations, and the required samples that are needed for achieving a certain numerical accuracy. For the first one, assuming a certain platform, one should consider the impact on the signal integrity/energy, i.e., the noise floor that the experimental measuring devices can achieve, and potential path length errors that might occur. For the second factor, if both our kernel and input are band-limited, one can possibly define a lower bound for the number of the waveguides. However, in principle, the number of waveguides depends very much on the nature of the mathematical problem to be solved.
 
\subsection*{Numerical Methods}
\subsubsection*{Rectangular integration}

The rectangular integration technique, also known as the midpoint rule, was used to evaluate the integral of Eq.~(\ref{eq1_IE}). This technique approximates the area under a curve as the sum of rectangles of different widths and heights. The integration domain $\left[a,b\right]$ was divided into $m$ subintervals of equal width $\Delta_v = (b-a)/m$.  The midpoint was extracted from each subinterval to build the vector ${\bf{v}}=[v_{1},v_{2},.....,v_{m}]$.  This represents the discretized version of the integration variable $v$. It was assumed that the input electromagnetic signal, $c(u)$, and the solution electromagnetic wave, $x(u)$, were also sampled at these same points so that $\bf{u}=\bf{v}$. Using this numerical approximation, Eq. (\ref{eq1_IE}) can be discretized by replacing the continuous variables $u$ and $v$  with the discrete variables $\bf{u}$ and $\bf{v}$, respectively, can be rewritten as
\begin{equation}\label{eq1_rect}
x\left( {{u_i}} \right) = c\left( {{u_i}} \right) + {\Delta _v}\sum\limits_{j = 1}^m {K\left( {{u_i},{v_j}} \right)x\left( {{v_j}} \right)} \quad \quad i = 1,.....,m
\end{equation}
where the integral has been replaced by a sum of areas of rectangles with width $\Delta_v $ and heights equal to the integrand value at midpoints. Eq.~(\ref{eq1_rect}) can then be expressed in matrix-vector form as ${\bf{x}}= {{\bf{c}}} + {\bf{K}}   {\bf{x}}$ where the term $\Delta_v $ has been incorporated in the kernel matrix $\bf{K}$. The latter has been interpreted as a  transmission matrix and implemented through an MZI mesh that evaluates numerically the integral of Eq.~(\ref{eq1_IE}). Other numerical integration techniques such as trapezoidal rule can be used to solve Eq.~(\ref{eq1_IE}) with the proposed platform. 

\subsubsection*{Finite-difference method}
The finite difference method is a widely used method for the solution of differential equations~\cite{leveque2007finite}. The main feature is that a continuous differential equation of the form 
\begin{equation}
\mathcal{D}
  \begin{cases}
    \frac{d^2f(x)}{dx^2}+p(x)\frac{df(x)}{dx}+q(x)f(x)=g(x), & x\in[x_d,x_u]\\
    f(x_d)=a,~f(x_u)=b
  \end{cases}
\end{equation}
result in the following difference equation
\begin{equation}
\mathcal{D}_\Delta:~
  \begin{cases}
    r(x_j)\left(f_{j+1}-2f_j+f_{j-1}\right)+p(x_j)\frac{f_{j+1}-f_{j-1}}{2}(\Delta x)+q(x_{j})f_j(\Delta x)^2=g(x_j)(\Delta x)^2, & j\in\{1,...,m\}\\
    f_0=a, f_{m+1}=b
  \end{cases}
\end{equation}
The discretization of the differential of first and second order can be either by applying central (CD), forward (FoD), and backward (BD) differences. As shown in the previous section the resulted difference matrix reads
\begin{equation}
  \vec{A}=\vec{R}(x) \vec{A}_{\Delta x}+\vec{P}(x) \vec{P}_{\Delta x}~\Delta x+\vec{Q}(x)~(\Delta x)^2
\end{equation}
In particular, the second difference matrix reads
\begin{equation}
  \vec{A}_{\Delta x}=
\left(
  \begin{array}{ccccc}
    -2 & 1 & 0 & \dots & 0 \\
    1 & -2 & 1 & \ddots & \vdots \\
    \vdots & \ddots & \ddots & \ddots & \vdots \\
    \vdots & \ddots & 1 & -2 & 1 \\
    0 & \dots & 0 & 1 & -2 \\
  \end{array}
  \right)
\end{equation}
where the first and last row implement the FoD and the BD scheme, respectively. This matrix is almost diagonally dominant with real and positive diagonal entries, hence all (or the majority) of the eigenvalues are positive. The matrix $\vec{R}(x)=\text{diag}\left(r(x_1),r(x_2),...,r(x_m)\right)$ is a diagonal matrix derived by the values of $r(x)$. Similarly, the first difference matrix reads
\begin{equation}
  \vec{P}_{\Delta x}=\frac{1}{2}
  {
\left(
  \begin{array}{ccccc}
    0 & 1 & 0 & \dots & 0 \\
    -1 & 0 & 1 & \ddots & \vdots \\
    \vdots & \ddots & \ddots & \ddots & \vdots \\
    \vdots & \ddots & -1 & 0 & 1 \\
    0 & \dots & 0 & -1 & 0 \\
  \end{array}
  \right)}
\end{equation}
where the same FoD/BD scheme is used and $\vec{P}(x)=\text{diag}\left(p(x_1),p(x_2),...,p(x_m)\right)$. The first difference matrix is skew-symmetric with real entries, implying that its eigenvalues are imaginary. The last matrix is a diagonal matrix $\vec{Q}(x)=\text{diag}\left(0,q(x_2),...,0\right)$ 

The input vector $\vec{b}$ for this particular Dirichlet-type boundary conditions reads
\begin{equation}
  \vec{c}=
  \left(
  \begin{array}{c}
    g(x_1)(\Delta x)^2-\left(1-p(x_1)\Delta x\right)a\\
    g(x_2)(\Delta x)^2\\
    \vdots\\
    g(x_{m-1})(\Delta x)^2\\
    g(x_{m})-\left(1+p(x_m)\Delta x\right)b
  \end{array}\right)
\end{equation}
Note that the input vector $\vec{c}$ changes depending on the boundary-type problem that is used. In our case the FoD/BD scheme was used as a way to increase the accuracy of the results at the boundaries. Assuming a much denser discretization (larger than the 11 points), then a uniform CD scheme could also be used without affecting the accuracy of the results. 

\subsection*{On the convergence of the iteration and GDM}

The optimal choice of the scaling factor $\alpha_\lambda$ is when it is proportional to the minimum and maximum eigenvalues of the operator, i.e.,   $\alpha_\lambda=\frac{2}{\lambda_1+\lambda_n}$, where $\lambda_1$ and $\lambda_n$ are the largest and smallest eigenvalues of the matrix. Knowledge of the operator's larger and smallest eigenvalues requires certainly some calculations. In order to avoid these calculations, one can utilize some of the eigenvalue inclusion theorems that are available, such as Schur's or Gershgorin theorems~\cite{ErwinKreyszig2011}. In the case where a different scaling factor is used the convergence will not be optimum, hence cannot be described by Eq.(\ref{eq:bound}). However, the convergence of the method to the desired level of accuracy can happen well before the bound is met even for optimum scaling factor. 

Finally, the iteration used above have been invented multiple times carrying a multitude of names such as Richardson's iteration~\cite{Richardson1911}, Landweber iteration~\cite{Landweber1951}, steepest or GDM~\cite{ShewchukJonathan1994}. An interesting feature with regards to its convergence should be mentioned here. Briefly, for a given positive semidefinite matrix $\vec{A}\in\mathbb{C}^{m\times m}$ we have that the iteration $x_{n+1}=\alpha\vec{c}+\left(\vec{I}-\alpha\vec{A}\right)x_n$ converges as long as $\vec{I}-\alpha\vec{A}$ has norm less than 1 and its actual convergence depends on how much this norm is less than 1. Let $\lambda_1\leq\lambda_2\leq...\leq \lambda_m$ are the eigenvalues of $\vec{A}$, then the norm of the kernel minimizes when $\alpha=\alpha_\lambda=\frac{2}{\lambda_1+\lambda_m}$. 

For this type of scaling value the $\vec{x}_n$ converges to the desired solution $\vec{x}$ when the norm is less than the desired accuracy level $\varepsilon$, i.e, after $n$ iterations we have that the value is bounded by  
\begin{equation*}
||\vec{x}-\vec{x}^{n}||=||\left(\vec{I}-\alpha_\lambda\vec{A}\right)^n\vec{x}||\leq||\left(\vec{I}-\alpha_\lambda\vec{A}\right)^n||~||\vec{x}||=\left(1-\alpha_\lambda\right)^n||\vec{x}||\leq e^{-\alpha_\lambda n}||\vec{x}||
\end{equation*}
and for $\frac{||\vec{x}-\vec{x}^{n}||}{||\vec{x}||}\leq\varepsilon$ we finally have that the iteration will converge after 
\begin{equation}
    n=\frac{1}{\alpha_\lambda}\ln{\left(1/\varepsilon\right)}=\frac{1}{2}\left(\kappa\left(\vec{A}\right)+1\right)\ln{\left(1/\varepsilon\right)}
\end{equation}
iterations. Assuming that each iteration requires time $t$ then we reach Eq.~(\ref{eq:bound}). For the case of GDM where we use the normal matrix $\vec{A^*A}$ then the condition number is modified accordingly. 

As a side note, in realistic implementations (assuming noise and other intrinsic system imperfections) the kernel matrix is followed by $(\vec{I}-\vec{A^*A} + \delta\vec{N})$, where $\vec{N}$ is a random noise matrix while $\delta$ is a small pertrubation. In the ideal case we have $\delta\to0$. Assuming a certain stochastic dependecne of the noice matrix the above generic quantity could be simplified to be $(\vec{I}-\vec{A^*A} +\delta\vec{I})$, implying that the Kernel represents a Tikhonov regularization scheme~\cite{Kleinman1992,Jensen2011}, hence the quantity $(\vec{A^*A})^{-1}$ has always an answer. 

\noindent\textbf{Acknowledgements}\\
This work was supported in part by the US Air Force Office of Scientific Research (AFOSR) Multidisciplinary University Research Initiative (MURI) grant number FA9550-17-1-0002.

\noindent\textbf{Author contributions}\\
N.E. initiated the research, acquired the funds, and presented the idea for solving mathematical equations with the Miller architecture. The DCM architecture was proposed by N.E., developed by D.C.T, and further refined by D.C.T. and B.E. D.C.T. performed all system level simulations. M.J.M. implemented the Miller algorithm and M.J.M. and B.E. developed the code for the IE solution. D.C.T. developed the code for the DE problems and extended the usage of metadevices for inverting general rectangular matrices. All authors discussed and interpreted all the results reported in this manuscript.  D.C.T. wrote the first draft of the manuscript and all authors discussed, commented, and edited the manuscript. \\

\noindent\textbf{Competing interests}\\
A non-provisional patent application has been filed on some of the ideas reported here, and more patent applications on other aspects of this work may be filed in the near future. N.E. is a strategic scientific advisor/consultant to Meta Materials, Inc.  No other conflict of interest is present.  \\

\noindent\textbf{Data availability}\\
All data (simulation files, raw data, figures, etc.) are available by the corresponding authors upon reasonable request.\\

\noindent\textbf{Code availability}\\
All codes (Matlab scripts etc.) are available by the corresponding authors upon reasonable request.


\bibliography{main}

\begin{thebibliography}{10}
\bibitem{supp2}
\bibinfo{title}{This holds for strictly diagonal matrices in our case, however,
  the introduction of a small positive perturbation allows us to expand this
  argument.}
\end{thebibliography}

\end{document}